\documentclass[aps,pre,twocolumn,showpacs,groupedaddress,footinbib]{revtex4-1}
\usepackage{amsmath,amsfonts,amssymb}
\usepackage{color}
\usepackage{graphicx}
\usepackage{subfig}

\begin{document}


\title{Multi-Dimensional Elephant Random Walk with Coupled Memory}


\author{Vitor M. Marquioni}
\email[]{vimarqmon@gmail.com}
\affiliation{Instituto de F\'isica de S\~ao Carlos, Universidade de S\~ao Paulo, CP369, 13560-970, S\~ao Carlos, SP, Brazil}


\date{\today}

\begin{abstract}
The elephant random walk (ERW) is a microscopic, one-dimensional, discrete-time, non-Markovian random walk, which can lead to anomalous diffusion due to memory effects.
In this study, I propose a multi-dimensional generalization in which the probability of taking a step in a certain direction depends on the previous steps in other directions. The original model is generalized in a straightforward manner by introducing coefficients that couple the probability of moving in one direction with the previous steps in all directions.
I motivate the model by first introducing a two-elephant system and then elucidating it with a specific coupling. With the explicit calculation of the first moments, I show the existence of two newsworthy relative movement behaviours: one in which one elephant follows the other, and another in which they go in opposite directions. With the aid of a Fokker-Planck equation, the second moment is evaluated and two new super-diffusion regimes appear, not found in other ERWs. Then, I re-interpret the equations as a bidimensional elephant random walk model, and further generalize it to $N-$dimensions.
I argue that the introduction of coupling coefficients is a way of extending any one-dimensional ERW to many dimensions.

\end{abstract}

\pacs{05.40.Fb, 02.50.Ey, 05.10.Gg}

\maketitle

\section{Introduction}
The term Random Walk (RW) was coined by the statistician Karl Pearson in a brief letter to Nature in 1905 \cite{pearson1905problem} and is now an important and widely used tool in the modeling of a myriad of stochastic processes found in biological systems, finance and physical processes \cite{einstein1905brownian,chandrasekhar1943stochastic,turchin1998quantitative,bachelier1900bachelier}. The simplest kind of RW considers no dependence of the future steps on the previous ones of the walk. This absence of \emph{memory} is the Markovian property, i.e., a process is called Markovian if the probability of the next state depends only on the present state. However, in some cases of interest, such as the growth of a polymer diluted in appropriate solvents \cite{hughes1995random}, neglecting the past is a remarkable mistake, since a growing molecule cannot polymerize over itself. Self-avoiding walks appeared in this context as a non-Markovian process that prohibits visiting a site more than once \cite{hughes1995random}.
In the last decade, a new class of non-Markovian random walks was introduced and studied.
In these systems, the walker has a memory mechanism so that the probability of taking a step forward or backward depends somehow on the previous steps. This differs from the self-avoiding walk because the past movements do not forbid a site to be revisited but influence the decision about the next step.

The first microscopic RW of this new class, which is referred to as \emph{Elephant Random Walks}, ERW, was proposed and analytically solved by Sch\"utz and Trimper in 2004 \cite{schutz2004elephants}. In this model, a one-dimensional elephant takes steps to the right or to the left and the probability of each step depends on the whole history of the elephant. 
The memory effect is due to a single parameter $p \in [0,1]$ that can lead to differing diffusion behavior. Here, $p$ is the probability of mimicking a randomly chosen previous step. If $p<1/2$, the walker's mean displacement tends to zero as the number of steps increases, while it tends to infinity if $p>1/2$. The latter is said to be a \emph{traditionalist} elephant and the former a \emph{reformer} one. For $p=1/2$, the walk is Markovian and it equates a simple random walk. If $p<3/4$, the walk is normal-diffusive and if $p>3/4$, the walk is super-diffusive.

Many variations of the ERW have been proposed \cite{cressoni2007amnesticallyalzheymer,kumar2010memory,cressoni2013exact2d,kim2014anomalous,harbola2014memory,di2017elephantquantum}, and a lot of mathematical results \cite{dasilva2013nongaussian,coletti2016central,coletti2017strong,bercu2017multi} and physical properties \cite{dasilva2006discrete,dasilva2008spontaneous,kenkre2007analytic,ferreira2010anomalous} were found in this kind of non-Markovian random walk.
In this paper, I present a multi-dimensional extension of the ERW in which the probability of taking a step in any direction might depend on the previous steps taken in any direction.
By introducing coupling coefficients, the model makes the walker look back to all directions in order to decide to move forward or backward in each dimension at the next step, taking one unit step per direction, per time unit step. 
This model can also be understood as a stochastic rule to walk a single step of size $d^{1/2}$ in a suitably chosen $d-$dimensional hyper-cubic lattice. The walker in this model is also similar to the original elephant, but it remembers not only its own steps, but also the steps of another elephant, and as such the next step will take into account all walking histories. Under this perspective, two different regimes from two memory-coupled individuals emerge: a chasing behaviour and a distancing one.
In addition, the lowest-order Fokker-Planck equation equivalent to this process is computed by means of the usual continuous limit approach in order to present a continuous analogue which exhibits the same diffusion behavior found in the discrete model, besides the two regimes listed above. Also, this continuous approach allows for the calculation of the second moment, which is difficult to be done in the discrete case. Two new anomalous diffusion behavioral patterns appear.

In section II, I introduce a two-elephant model, motivated by the question \emph{What would happen if an elephant could also remember the past of another elephant?}. Here, the coupling coefficients appear as a probability distribution of an elephant choosing to mimic its own past or the past of the other elephant. I explicitly show a recursion formula for the first moment and how to calculate recursion formulas for the higher-order moments. In section III, the model is elucidated with a specific coupling, called the \emph{Cow-and-Ox Model}, a coupling that allows only one walker (the Ox) to remember the past of the other (the Cow). This situation is introduced in \cite{moura2017randomautism} in order to derive a model to quantify autism, but not with the same formalism or analytical results. The first moment evaluation comes up with two noteworthy behavioral patterns: one in which one walker chases the other, and another in which the walkers separate from each other (both regimes on average). I also show that super-diffusion can still occur for a specific parameters regime. A continuous limit approach is presented in section IV, where the two elephants now represent two different directions of a single elephant. Considering a walk on a square lattice, the jump equation is calculated by introducing the complex characteristic functions of two directions, and then it is approximated to a two-dimensional Fokker-Planck equation, which is an approximation to a long time limit. This equation is usefull to calculate the second moment, which exhibits two new anomalous diffusions: one faster marginally super-diffusive and one super-diffusive slightly faster than those found in the original ERW. I conclude by extending the two-elephant model to an $N-$dimensional one, giving appropriate rules to the multi-dimensional steps. Evidently, the $N-$dimensional model can also be interpreted as a model of $N$ interacting elephants.

\section{The two-elephant model}\label{section2}

In the original ERW model, the position of the walker at time $t+1$, $X_{t+1}$, is given by
\begin{equation}
X_{t+1}=X_t+\sigma_{t+1} \label{erw}
\end{equation}
where $\sigma_{t+1}=\pm1$ with probabilities given as follows:
\begin{enumerate}
	\item at time $t+1$, a time $t'$ is chosen randomly from the set $\{1,\ldots,t\}$ with uniform probability $1/t$;
	\item $p$ is the probability of following the step taken at $t'$, i.e., with probability $p$, $\sigma_{t+1}=\sigma_{t'}$, and with probability $1-p$, $\sigma_{t+1}=-\sigma_{t'}$. This probability can be written as
	\begin{equation}
	\mathcal{P}\left[\sigma_{t+1}=\pm\sigma_{t'}\right]=\dfrac{1}{2}\left[1+(2p-1)\sigma_{t+1}\sigma_{t'}\right];\label{condprob1}
	\end{equation}
	\item at the first step, $\sigma_1=+1$ with probability $q$ and $\sigma_1=-1$ with probability $1-q$, i.e.,
	\begin{equation}
	\mathcal{P}\left[\sigma_{1}=\pm1\right]=\dfrac{1}{2}\left[1+(2q-1)\sigma_{1}\right].
	\end{equation}
\end{enumerate}

Now, two elephants are assumed to walk on the same line. The first elephant, whose position will be denoted by $X_t^1$, might remember not only its own past, but the whole past of the other elephant as well, $X_t^2$. Then, the random walk can be constructed as
\begin{equation}
X_{t+1}^i=X_t^i+\sigma_{t+1}^i,
\end{equation}
with $i=1,2$.
The rules of this RW are the following:
\begin{enumerate}
	\item at time $t+1$, elephant $i$ chooses an elephant $k=1,2$ with probability $\gamma_k^i$, ($\gamma_1^i+\gamma_2^i=1$);
	\item then, a time $t'$ is randomly chosen from the set $\{1,\ldots,t\}$ with uniform probability;
	\item now, the step $\sigma_{t+1}^i$ of elephant $i$ will be
		\begin{equation}
			\sigma_{t+1}^i=\left\{
			\begin{aligned}
			&+\sigma_{t'}^k,\hspace{5mm}\textrm{with probability}\hspace{5mm}p_k^i;\\
			&-\sigma_{t'}^k,\hspace{5mm}\textrm{with probability}\hspace{5mm}1-p_k^i;
			\end{aligned}\right.\nonumber
		\end{equation}
		i. e.,
		\begin{equation}
			\mathcal{P}\left[\sigma_{t+1}^i=\pm\sigma_{t'}^k|\sigma_{t'}^k\right]=\dfrac{1}{2}\left[1+(2p_k^i-1)\sigma_{t+1}^i\sigma_{t'}^k\right];
		\end{equation}
	\item lastly, the first step is taken with probability
		\begin{equation}
			\mathcal{P}\left[\sigma_{1}^i=\pm1|\textrm{direction }k\right]=\dfrac{1}{2}\left[1+(2q_k^i-1)\sigma_{1}^i\right];
		\end{equation}
		in which I choose the dependence on $k$ simply to keep a simetric notation.
\end{enumerate}

With these new rules, the probability of the step $\sigma_{t+1}^i=\sigma$, given the chosen steps $\{\sigma_{t'}^1,\sigma_{t'}^2\}$, is
\begin{equation}
	\mathcal{P}\left[\sigma_{t+1}^i=\sigma|\sigma_{t'}^{1,2}\right]=\sum_{k=1}^{2}\dfrac{1}{2}\left[1+(2p_k^i-1)\sigma\sigma_{t'}^k\right]\gamma_k^i.\label{condprob2}
\end{equation}
Comparing this equation with Eq.(\ref{condprob1}), $\gamma_k^i$ can be viewed as the coupling coefficient of elephant $i$ on elephant $k$.
In addition, the first step is given with probability
\begin{equation}
	\mathcal{P}\left[\sigma_{1}^i=\sigma\right]=\sum_{k=1}^{2}\dfrac{1}{2}\left[1+(2q_k^i-1)\sigma\right]\gamma_k^i.
\end{equation}
By using Eq.(\ref{condprob2}), one can calculate the conditional probability $\mathcal{P}\left[\sigma_{t+1}^i=\sigma|\sigma_1^1,\ldots,\sigma_t^1;\sigma_1^2,\ldots,\sigma_t^2\right]$ as being
\begin{equation}
\mathcal{P}\left[\sigma_{t+1}^i=\sigma|\{\sigma_{1,\ldots,t}^{1,2}\}\right]=\dfrac{1}{2}+\sigma\sum_{k=1}^2\dfrac{x_t^k\alpha_k^i\gamma_k^i}{2t},\label{eq9}
\end{equation}
with $\alpha_k^i=2p_k^i-1$ and $x_t^k=X_t^k-X_0^k$ being the displacement of elephant $i$. The conditional mean increment of each elephant is
\begin{align}
\left\langle\sigma_{t+1}^i=\sigma|\{\sigma_{1,\ldots,t}^{1,2}\}\right\rangle&=\sum_{\sigma=\pm1}\sigma\mathcal{P}\left[\sigma_{t+1}^i=\sigma|\{\sigma_{1,\ldots,t}^{1,2}\}\right]\nonumber\\
&=\sum_{k=1}^2\dfrac{x_t^k\alpha_k^i\gamma_k^i}{t}\label{increm}
\end{align}

The application of Eq.(\ref{increm}) results in the recursion formula for the first moment of each displacement
\begin{equation}
\left\langle x_{t+1}^i\right\rangle=\sum_{k=1}^2\left(\delta_{ki}+\dfrac{\gamma_k^i\alpha_k^i}{t}\right)\left\langle x_{t}^k\right\rangle\label{eq11}
\end{equation}
and, by defining another shifted parameter $\beta_k^i=2q_k^i-1$, one can get
\begin{equation}
\left\langle x_{1}^i\right\rangle=\sum_{k=1}^2\beta_k^i\gamma_k^i\label{beta}
\end{equation}

For higher-order displacement moments, we take
\begin{equation}
	\prod_{j=1}^{n}x_{t+1}^{i_j}=\prod_{j=1}^{n}\left(x_t^{i_j}+\sigma_{t+1}^{i_j}\right),\label{highmom}
\end{equation}
where $i_j=1,2$ and $n$ is the order of the moment considered. Then, we first take the conditional average given an specific history of both elephants, and finally take the average over all possible histories. In general, the recursion relations to higher-order moments are all of the form
\begin{equation}
\mathbb{M}_{t+1}=\mathbb{H}_t+\mathbb{G}_t\mathbb{M}_t\label{eq14}
\end{equation}
where $\mathbb{M}_t$ and $\mathbb{H}_t$ are column matrices, $\mathbb{M}_t$ represents the $n$-th moment matrix, and $\mathbb{G}_t$ is a square matrix. The number of entries of $\mathbb{M}_t$ matrix is the number of moments of the considered order, given by $n+1$.
A solution can be encountered from the following formula
\begin{equation}
\mathbb{M}_t=\left(\prod_{k=t-1}^1\mathbb{G}_k\right)\mathbb{M}_1+\sum_{i=1}^{t-2}\left(\prod_{k=t-1}^{i+1}\mathbb{G}_k\right)\mathbb{H}_i+\mathbb{H}_{t-1}.\label{rec}
\end{equation}

\section{The Cow-and-Ox Model}\label{section3}

In order to clarify the model I have introduced, now I present the case in which the first elephant (said to be the \emph{Cow}) does not depend on the second (said to be the\emph{Ox}), but the Ox depends on the Cow, so the coupling coefficients are provided as follows
\begin{equation}
\left\{
\begin{array}{l}
\gamma_1^1=1;\\
\\\gamma_2^1=0;\\
\\\gamma_1^2=\gamma\ne0;\\
\\\gamma_2^2=1-\gamma.
\end{array}
\right.\
\end{equation}
and so one can calculate the first moment
\begin{equation}
\begin{pmatrix}
\langle x_{t+1}^1\rangle \\
\\
\langle x_{t+1}^2\rangle
\end{pmatrix}
=\dfrac{1}{t}
\begin{pmatrix}
t+\alpha_1^1 & 0 \\
\\
\gamma\alpha_1^2 & t+(1-\gamma)\alpha_2^2
\end{pmatrix}
\begin{pmatrix}
\langle x_t^1\rangle \\
\\
\langle x_t^2\rangle
\end{pmatrix}
\end{equation}
which has the solution
\begin{widetext}
\begin{equation}
\left\langle x_t^1\right\rangle=\dfrac{\Gamma(t+\alpha_1^1)}{\Gamma(t)\Gamma(\alpha_1^1+1)}\left\langle x_1^1\right\rangle,\label{sol2}
\end{equation}
\begin{align}
\left\langle x_t^2\right\rangle&=\gamma\alpha_1^2\left\{\dfrac{\Gamma(t-1+\alpha_1^1)}{\Gamma(t)\Gamma(\alpha_1^1+1)}+\dfrac{\Gamma(t+(1-\gamma)\alpha_2^2)}{\Gamma(t)\Gamma((1-\gamma)\alpha_2^2+2)}\right.
+\left.\dfrac{\Gamma(t+(1-\gamma)\alpha_2^2)}{\Gamma(t)\Gamma(\alpha_1^1+1)}\sum_{k=1}^{t-3}\dfrac{\Gamma(t-k-1+\alpha_1^1)}{\Gamma(t-k+(1-\gamma)\alpha_2^2))}\right\}\left\langle x_1^1\right\rangle\nonumber\\&+\dfrac{\Gamma(t+(1-\gamma)\alpha_2^2)}{\Gamma(t)\Gamma((1-\gamma)\alpha_2^2+1)}\left\langle x_1^2\right\rangle,\label{sol}
\end{align}
\end{widetext}
and from Eq.(\ref{beta}),
\begin{equation}
\left\langle x_1^1\right\rangle=\beta_1^1,
\end{equation}
\begin{equation}
\left\langle x_1^2\right\rangle=\gamma(\beta_1^2-\beta_2^2)+\beta_2^2.
\end{equation}

The asymptotic behaviour ($t\gg1$) of the solution is
\begin{equation}
\left\langle x_t^1\right\rangle\sim\dfrac{t^{\alpha_1^1}}{\Gamma(\alpha_1^1+1)}\left\langle x_1^1\right\rangle,\label{sol3}
\end{equation}
and for $\alpha_1^1\ne(1-\gamma)\alpha_2^2$,
\begin{widetext}
\begin{align}
\left\langle x_t^2\right\rangle&\sim\left[\dfrac{\gamma\alpha_1^2\left\langle x_1^1\right\rangle}{\Gamma(\alpha_1^1+1)(\alpha_1^1-(1-\gamma)\alpha_2^2)}\right] t^{\alpha_1^1}\nonumber\\&+t^{(1-\gamma)\alpha_2^2}\left[\dfrac{\left\langle x_1^2\right\rangle}{\Gamma(1+(1-\gamma)\alpha_2^2)}\right.+\left.\gamma\alpha_1^2\langle x_1^1\rangle\left(\dfrac{1}{\Gamma(2+(1-\gamma)\alpha_2^2)}\right.\right.-\left.\left.\dfrac{\Gamma(2+\alpha_1^1)/\Gamma(2+(1-\gamma)\alpha_2^2)}{(\alpha_1^1-(1-\gamma)\alpha_2^2)\Gamma(\alpha_1^1+1)}\right)\right].\label{sol4}
\end{align}
\end{widetext}

The behavior of the Cow is elephant-like, as expected, because it is not dependent on the Ox. The Cow is a \emph{reformer elephant} if $\alpha_1^1<0$ and it is a \emph{tradionalist elephant} if $\alpha_1^1>0$.

However, to complete the asymptotic behaviour analysis of the Ox, we need to compare the exponents $\alpha_1^1$ and $\left(1-\gamma\right)\alpha_2^2$. Three different behavioral patterns are possible: the first one is when
\begin{equation}
(i)\hspace{5mm}\alpha_1^1>(1-\gamma)\alpha_2^2\label{regimei}
\end{equation}
i. e., the probability of the Cow following its own past, is greater than the \emph{importance} that the Ox gives in following its own past. Under this condition, the asymptotic solution becomes
\begin{equation}
\left\langle x_t^2\right\rangle\sim\dfrac{\gamma\alpha_1^2\left\langle x_t^1\right\rangle}{\alpha_1^1-(1-\gamma)\alpha_2^2}\label{firstregime}
\end{equation}
so that on average the Ox and the Cow behave in the same way. Thus there are four regimes: $(i.a)$ $\alpha_1^2>0$ and $\alpha_1^2<\alpha_1^1-(1-\gamma)\alpha_2^2$ in which the Ox behaves \emph{like a detective}, it follows the Cow while always staying some steps behind \footnote{This regime gives rise to the characters of this model based on a traditional Brazilian song that says \emph{``where the Cow goes the Ox follows behind''}}; $(i.b)$ $\alpha_1^2>0$ and $\alpha_1^2>\alpha_1^1-(1-\gamma)\alpha_2^2$, in which the Ox goes the same direction as the Cow, but some steps ahead; $(i.c)$ $\alpha_1^2<0$, in which the Ox and the Cow go on opposite directions; $(i.d)$ $\alpha_1^2=0$,in which $\left\langle x_t^2\right\rangle\sim t^{(1-\gamma)\alpha_2^2}$, as can be seen from Eq.(\ref{sol}), which means that the Ox behaves independently. This happens because it has a Markovian dependence on the Cow, as in the limit case $p=1/2$ in the ERW \cite{schutz2004elephants}. In other words, the Ox has no dependence on the Cow.

The second behavioral pattern is provided by
\begin{equation}
(ii)\hspace{5mm}\alpha_1^1<(1-\gamma)\alpha_2^2\label{regimeii}
\end{equation}
then
\begin{widetext}
\begin{align}
\left\langle x_t^2\right\rangle\sim t^{(1-\gamma)\alpha_2^2}\left[\dfrac{\left\langle x_1^2\right\rangle}{\Gamma(1+(1-\gamma)\alpha_2^2)}\right.+\left.\gamma\alpha_1^2\langle x_1^1\rangle\left(\dfrac{1}{\Gamma(2+(1-\gamma)\alpha_2^2)}\right.\right.-\left.\left.\dfrac{\Gamma(2+\alpha_1^1)/\Gamma(2+(1-\gamma)\alpha_2^2)}{(\alpha_1^1-(1-\gamma)\alpha_2^2)\Gamma(\alpha_1^1+1)}\right)\right].
\end{align}
\end{widetext}
in which the Ox is almost fully decoupled from the Cow. In this case there are three regimes: if $
(ii.a)$ the expression between the brackets is positive, the mean displacement is greater than zero; if $
(ii.b)$, then the expression between the brackets is negative, the mean displacement is also negative, and if $
(ii.c)$, the expression between the brackets equals zero, we need to analize the coefficients of $t^{\alpha_1^1}$, as can be seen in Eq.(\ref{sol4}).

The third behavior happens when
\begin{equation}
(iii)\hspace{5mm}\alpha_1^1=(1-\gamma)\alpha_2^2
\end{equation}
whose asymptotic analysis cannot be made in Eq.(\ref{sol4}). However, by applying Eq.(\ref{sol}) one arrives at
\begin{equation}
\left\langle x_t^2\right\rangle\sim\gamma\alpha_1^2\ln (t)\langle x_t^1\rangle\label{thirdregime}
\end{equation}
in which the Ox marginally does the same as the Cow if $(iii.a)$ $(\alpha_1^2>0)$, or it distances itself from the Cow if $(iii.b)$ $(\alpha_1^2<0)$.

\begin{figure*}[htb]
	\includegraphics[scale=0.61]{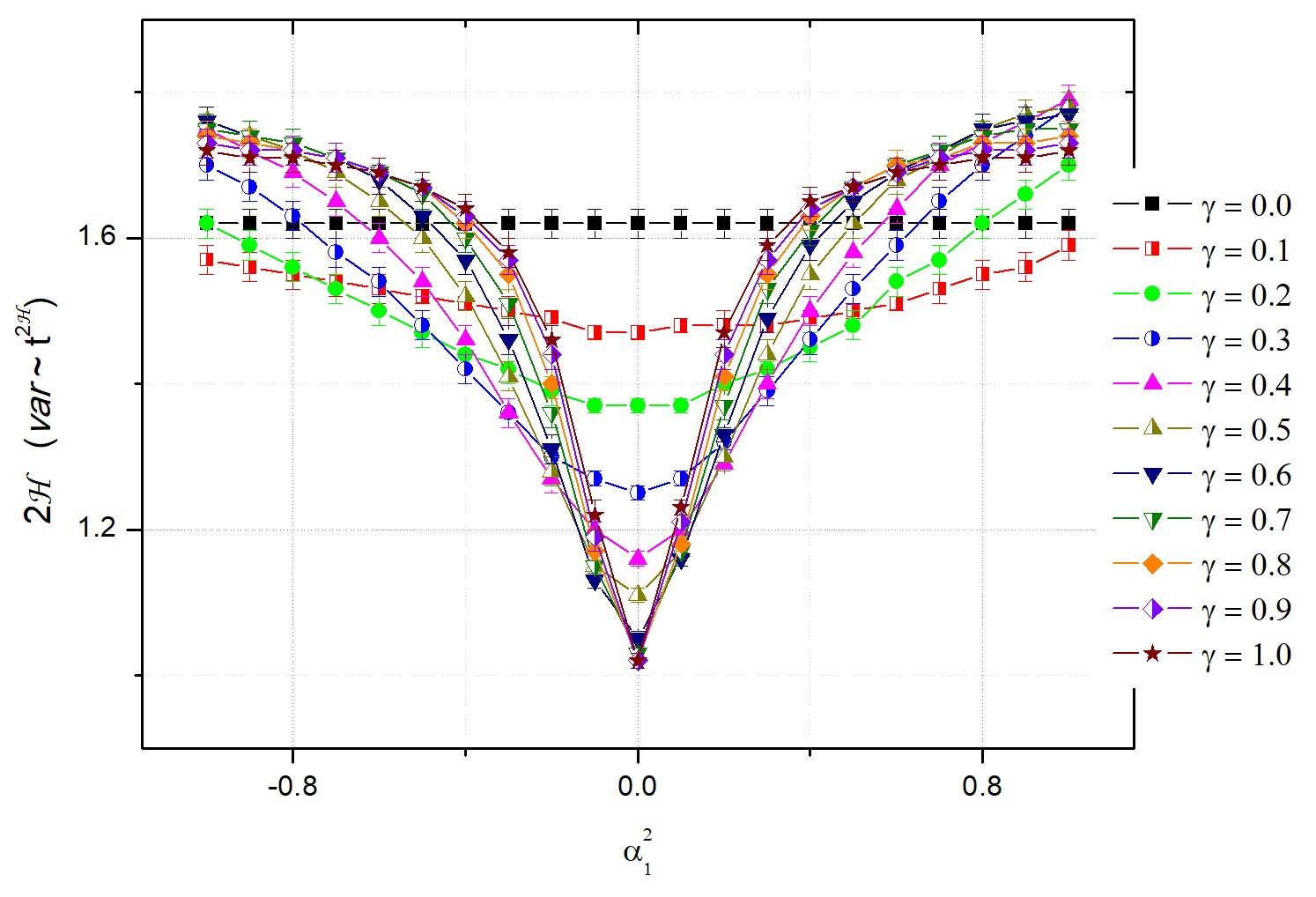}
	\caption{\label{figura1}(Color Online) Ox Diffusion Dependence on $\alpha_1^2$ and $\gamma$. Defining the Hurst exponent $\mathcal{H}$ as $\langle(x_t^i)^2\rangle-\langle x_t^i\rangle^2\sim t^{2\mathcal{H}}$, normal diffusion is defined as $2\mathcal{H}=1$, so that super-difusion happens when $2\mathcal{H}>1$. In this picture, I simulate the Ox diffusion behavior to different values of $\alpha_1^2$ and $\gamma$ for $10^3$ Ox walking $10^3$ steps each one. Here, $\alpha_1^1=\alpha_2^2=0.8$, $\alpha_2^1=0$ and $\beta_i^j=1.0$, to $i,j=1,2$.}
\end{figure*}

For the second moment,
\[
\mathbb{M}_t=\begin{pmatrix}
\langle\left(x_t^1\right)^2\rangle & \langle x_t^1x_t^2\rangle & \langle\left(x_t^2\right)^2\rangle
\end{pmatrix}^T
\]
according to equation (\ref{highmom}), and following the procedure described subsequently, one arrives at
\begin{widetext}
\begin{equation}
\mathbb{G}_t=
\begin{pmatrix}
1+2\dfrac{\alpha_1^1\gamma_1^1}{t} & 0 & 0 \\
\dfrac{\alpha_1^2\gamma_1^2}{t}+\dfrac{\alpha_1^1\gamma_1^1\alpha_1^2\gamma_1^2}{t^2} & 1+\dfrac{\left(\alpha_1^1\gamma_1^1+\alpha_2^2\gamma_2^2\right)}{t}+\dfrac{\alpha_1^1\gamma_1^1\alpha_2^2\gamma_2^2}{t^2} & 0 \\
0 & \dfrac{2\alpha_1^2\gamma_1^2}{t} & 1+2\dfrac{\alpha_2^2\gamma_2^2}{t}
\end{pmatrix}
\hspace{5mm}\textrm{and}\hspace{5mm}
\mathbb{H}_t=\begin{pmatrix}
1\\
\\
0\\
\\
1
\end{pmatrix}
\end{equation}
\end{widetext}
with $\mathbb{M}_1=\begin{pmatrix}
1 & \langle x_1^1\rangle\langle x_1^2\rangle & 1
\end{pmatrix}^T$. In the case of $\alpha_1^2=0$,  the solution is trivial, since the coupling between the Cow and the Ox vanishes (the dependence is Markovian), $\mathbb{G}_t$ becomes diagonal and the diffusion behavior is elephant-like:
\begingroup
\allowdisplaybreaks
\begin{align}
\langle\left(x_t^1\right)^2\rangle&=\dfrac{t}{2\alpha_1^1\gamma_1^1-1}\left(\dfrac{\Gamma\left(t+2\alpha_1^1\gamma_1^1\right)}{\Gamma\left(t+1\right)\Gamma\left(2\alpha_1^1\gamma_1^1\right)}-1\right)\label{asim1},\\
\langle x_t^1x_t^2\rangle&=\langle x_t^1\rangle\langle x_t^2\rangle,\\
\langle\left(x_t^2\right)^2\rangle&=\dfrac{t}{2\alpha_2^2\gamma_2^2-1}\left(\dfrac{\Gamma\left(t+2\alpha_2^2\gamma_2^2\right)}{\Gamma\left(t+1\right)\Gamma\left(2\alpha_2^2\gamma_2^2\right)}-1\right)\label{asim2},
\end{align}
\endgroup
and in the asymptotic limit,
\begin{equation}
\langle\left(x_t^i\right)^2\rangle\sim\left\{\begin{matrix}
\dfrac{t}{1-2\alpha_i^i\gamma_i^i},\hspace{5mm}\alpha_i^i\gamma_i^i<1/2\\
\\
t\ln t,\hspace{5mm}\alpha_i^i\gamma_i^i=1/2\\
\\
\dfrac{t^{2\alpha_i^i\gamma_i^i}}{\left(2\alpha_i^i\gamma_i^i-1\right)\Gamma\left(2\alpha_i^i\gamma_i^i\right)},\hspace{5mm}\alpha_i^i\gamma_i^i>1/2
\end{matrix}\right.\label{asymp}
\end{equation}
i.e., both walkers (the Ox and the Cow) might present super-diffusive behaviour if $\alpha_i^i\gamma_i^i>1/2$ when $\alpha_1^2=0$. If $\alpha_1^2\neq0$, the diffusion behaviour is not trivial, as can be seen in Fig.~\ref{figura1}.

\begin{figure*}[h]
	\begin{center}
		\subfloat[$\alpha_1^2=0$.]{
			\includegraphics[scale=0.5]{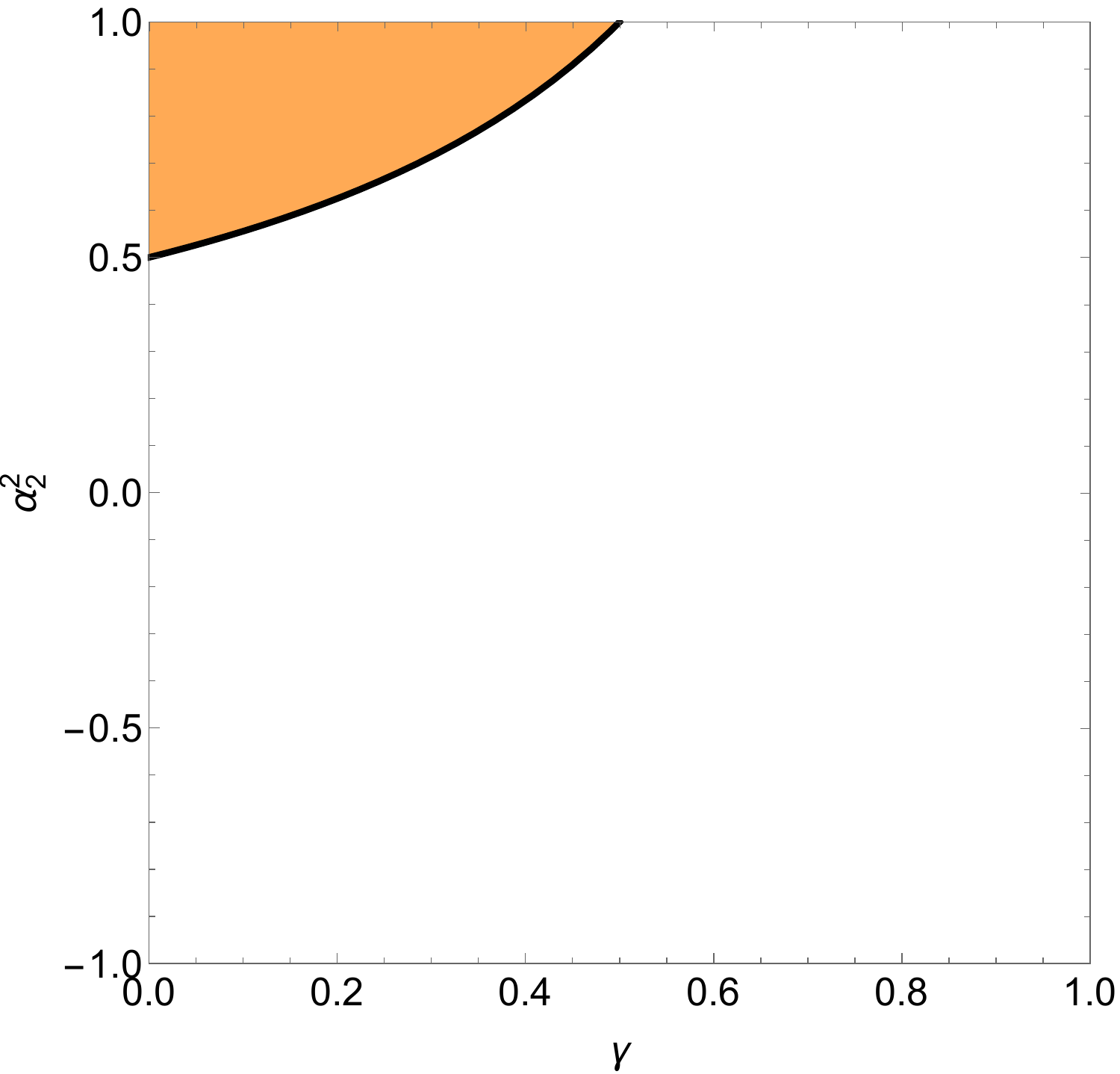}
			\label{figura2a}}
		\quad
		\subfloat[$\alpha_1^2\ne0$: first and second moment regions.]{
			\includegraphics[scale=0.5]{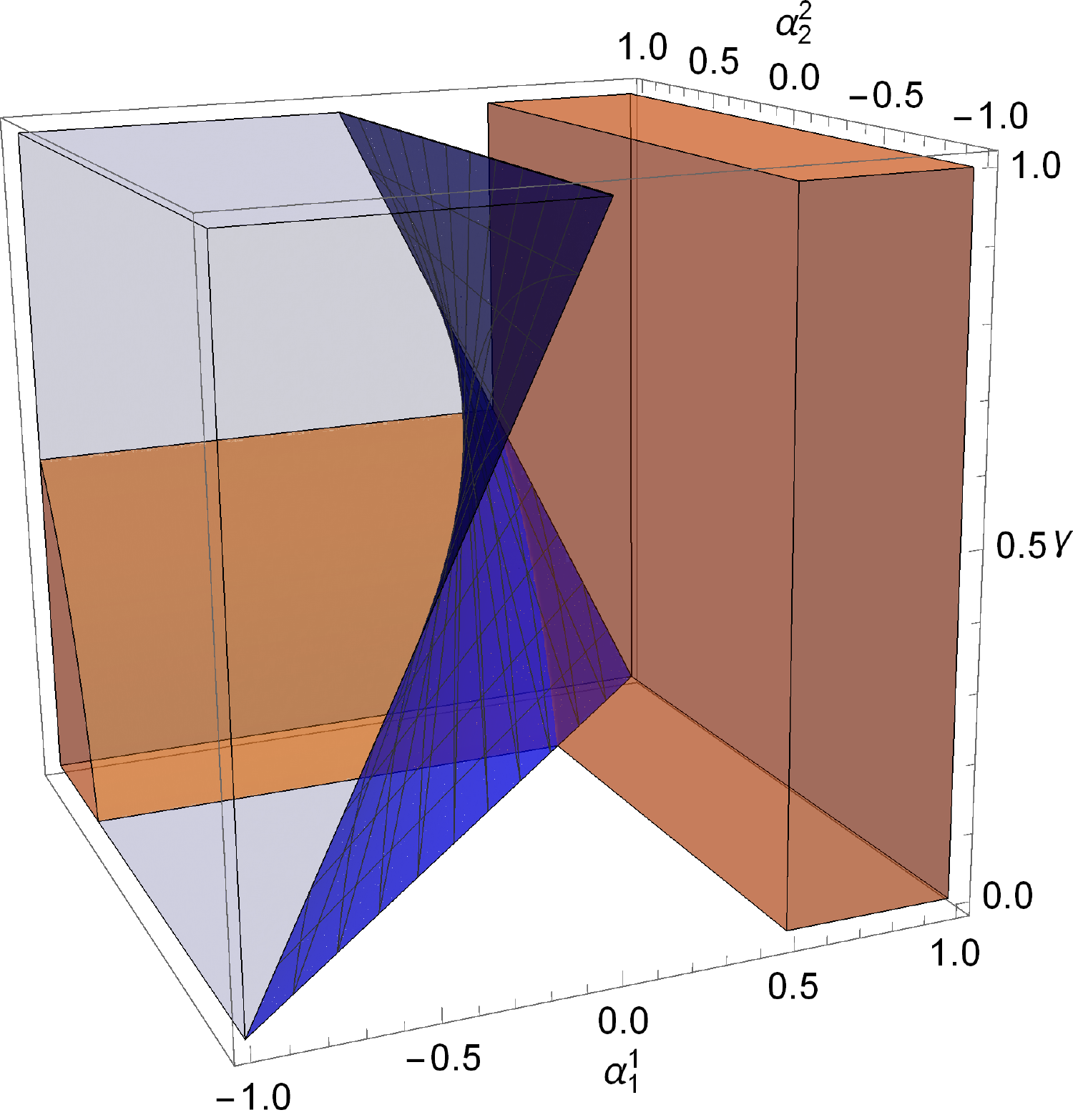}
			\label{figura2b}}
		\quad
		\subfloat[$\alpha_1^2\ne0$: marginally super-diffusive surfaces.]{
			\includegraphics[scale=0.5]{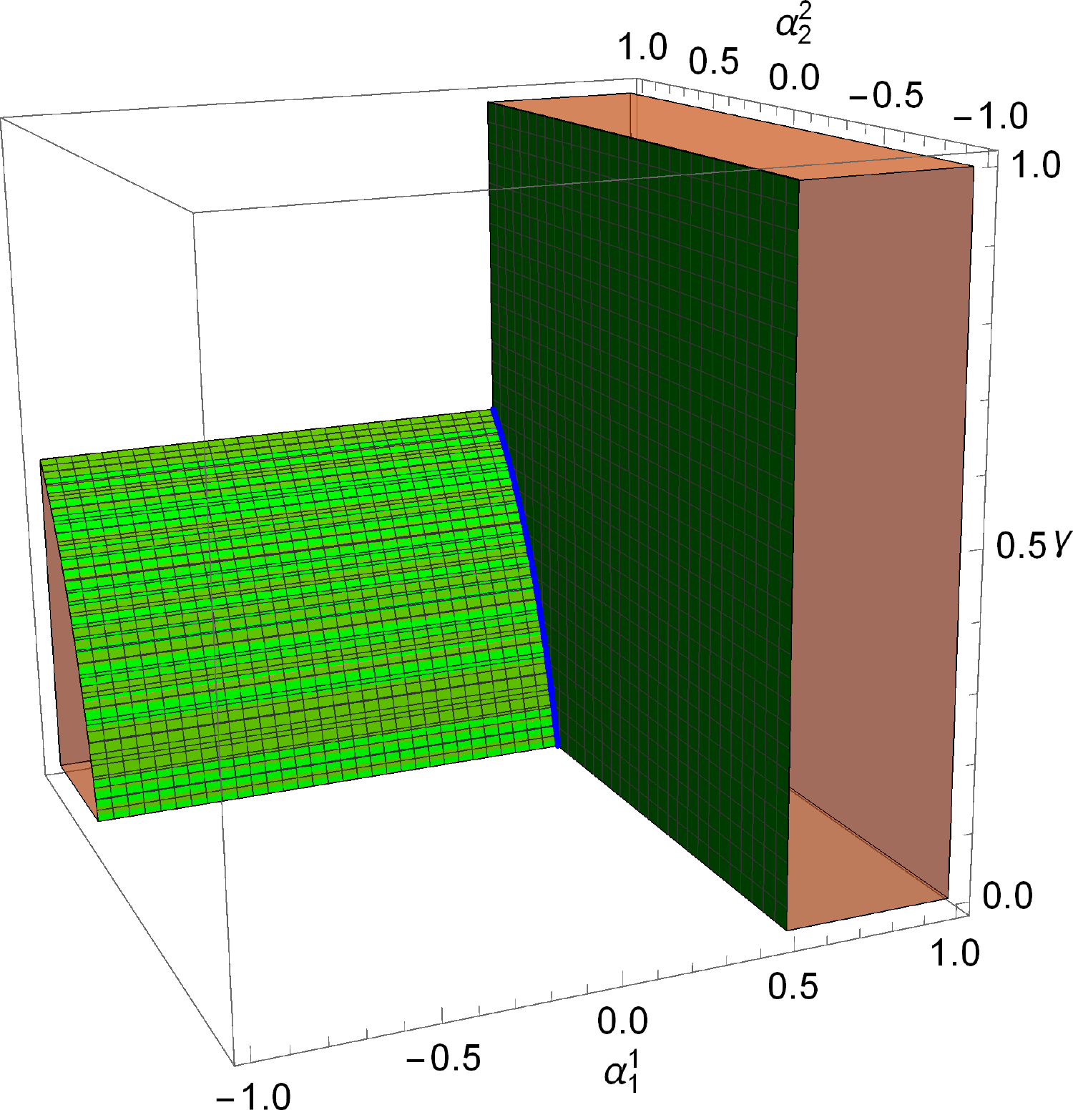}
			\label{figura2c}}
		\quad
		\subfloat[$\alpha_1^2\ne0$: $\alpha_1^1=(1-\gamma)\alpha_2^2>1/2$.]{
			\includegraphics[scale=0.5]{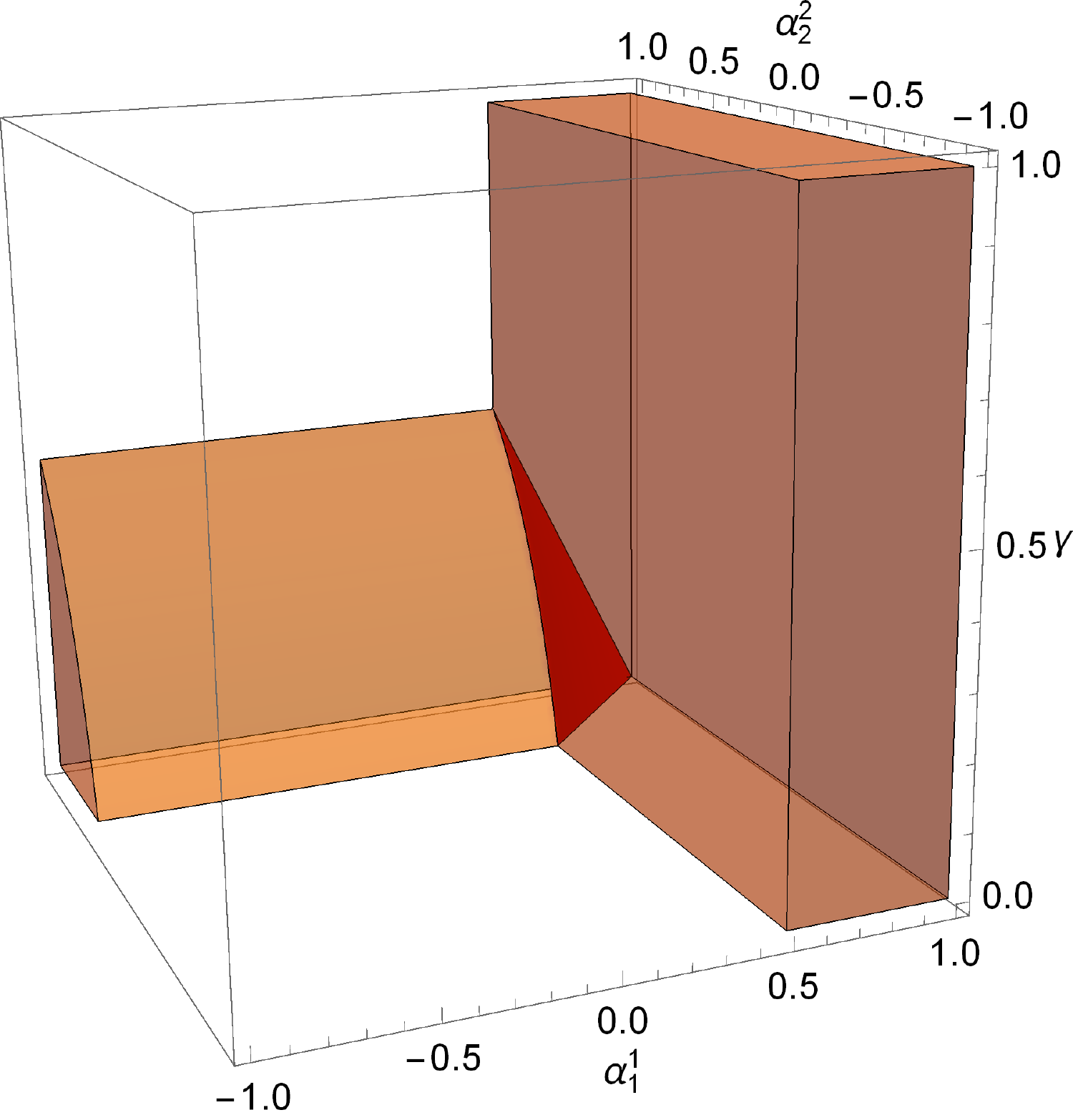}
			\label{figura2d}}
	\end{center}
	\caption[Ox Phase Diagram.]{(Color Online) Ox Phase Diagram. The diffusion is anomalous within the orange regions in all figures, and normal outside. In (\ref{figura2a}), because $\alpha_1^2=0$, the Ox has a Markovian dependence on the Cow, so its behaviour does not depend on $\alpha_1^1$. In the white region, $(1-\gamma)\alpha_2^2<1/2$, the Ox is normal diffusive; in the orange region, $(1-\gamma)\alpha_2^2>1/2$ and $\langle (x_t^2)^2\rangle\sim t^{2\gamma\alpha_2^2}$, then the Ox is super-difusive; at the black line $(1-\gamma)\alpha_2^2=1/2$, the Ox is marginally super-diffusive, with $\langle (x_t^2)^2\rangle\sim tln(t)$, (equation (\ref{asymp})). In (\ref{figura2b}), the blue surface is the surface $\alpha_1^1=\alpha_2^2(1-\gamma)$, where the first moment satisfies Eq.(\ref{thirdregime}). At the blue side, $\alpha_1^1<\alpha_2^2(1-\gamma)$ and $\langle x_t^2\rangle\sim t^{(1-\gamma)\alpha_2^2}$, then, where $(1-\gamma)\alpha_2^2>1/2$ (orange region inside the blue one) the Ox is super-diffusive with $\langle (x_t^2)^2\rangle\sim t^{2(1-\gamma)\alpha_2^2}$, and it is normal diffusive otherwise. At the other side of the surface, $\alpha_1^1>\alpha_2^2(1-\gamma)$, the Ox' first moment is linearly dependent on the Cow's (Eq.(\ref{firstregime})) and where $\alpha_1^1>1/2$, (orange region outside the blue one), the Ox is super-diffusive with $\langle (x_t^2)^2\rangle\sim t^{2\alpha_1^1}$ and normal diffusive otherwise. In (\ref{figura2c}), the diffusion is marginally super-diffusive of type $\langle (x_t^2)^2\rangle\sim tln(t)$ at the green surface but faster at the blue line $\alpha_1^1=(1-\gamma)\alpha_2^2=1/2$, in which $\langle (x_t^2)^2\rangle\sim t\left[ln(t)\right]^3$. In (\ref{figura2d}), the red surface satisfies $\alpha_1^1=(1-\gamma)\alpha_2^2>1/2$, and the diffusion at that is even faster, $\langle (x_t^2)^2\rangle\sim t^{2\alpha_1^1}\left[ln(t)\right]^2$.}\label{figura2}
\end{figure*}

However, as discussed in \cite{ferreira2010anomalous}, for a random walk lacking subdiffusion, if $\langle x_t\rangle\sim t^{\delta}$ and $\langle (x_t)^2\rangle\sim t^{2\mathcal{H}}$, then $\mathcal{H}=\delta$ when $\delta>1/2$, and $\mathcal{H}=1/2$ when $\delta<1/2$. For the Cow, we know it is true, since it behaves as an elephant. Assuming this conjecture also holds for the Ox, it is possible to calculate its  second moment, and hence the diffusion behavior from equation (\ref{sol4}). For instance, when condition (\ref{regimei}) is valid, both Cow and Ox diffuse the same way ($\langle (x_t^i)^2\rangle\sim t^{2\alpha_1^1}$), but when condition (\ref{regimeii}) is valid, the Ox can be super-diffusive whereas the Cow can remain normal diffusive. Moreover, it is worth noting that if the Cow is super-diffusive, the Ox will be too, regardless of the considered regime. These results (concerning the Ox movement) are disposed at the phase diagrams of Fig.~\ref{figura2}. The colored blocks represent different behavioral patterns regarding the first and second moment described so far. Nevertheless, the surfaces in the diagrams possess non-trivial diffusion behaviours, which will be calculated below. 

\section{Continuous Limit}\label{section4}

The process described so far also describes a bidimensional elephant walking on a flat surface in which each step taken in each direction depends on the history of both directions. As such, from now on, I shall treat both elephants as a unique bidimensional one, whose displacement is given by the vector $\overrightarrow{x_t}=\left(x_t^1,x_t^2\right)$, with steps $\overrightarrow{\sigma_t}=\left(\sigma_t^1,\sigma_t^2\right)$.

In order to introduce a continuous approximation to this bidimensional walk, I calculate a Fokker-Planck equation using the jumping process that can be found with the complex characteristic function of two directions $Q_t(\overrightarrow{k})=\langle e^{i\overrightarrow{k}\cdot\overrightarrow{x}_t}\rangle$ where $\overrightarrow{k}=k^1\hat{x}^1+k^2\hat{x}^2$ and $\overrightarrow{x}_t=x^1_t\hat{x}^1+x^2_t\hat{x}^2$, with $\hat{x}^{1,2}$ being the orthonormal basis of the square lattice where the walk is supposed to be, so that
\begin{align}
Q_{t+1}(\overrightarrow{k})&=\cos(k^1)\cos(k^2)Q_t(\overrightarrow{k})\nonumber\\&+\cos(k^1)\sin(k^2)\sum_{i=1}^2\dfrac{\partial Q_t}{\partial k^i}\dfrac{\alpha_i^2\gamma_i^2}{t}\nonumber\\&+\cos(k^2)\sin(k^1)\sum_{i=1}^2\dfrac{\partial Q_t}{\partial k^i}\dfrac{\alpha_i^1\gamma_i^1}{t}\nonumber\\&+\sin(k^1)\sin(k^2)\sum_{i=1}^2\sum_{j=1}^2\dfrac{\partial^2 Q_t}{\partial k^i\partial k^j}\dfrac{\alpha_i^1\gamma_i^1\alpha_j^2\gamma_j^2}{t^2}\label{reccomplex}
\end{align}
which can be solved with inverse Fourier transform, leading to the jumping process
\begin{align}
\mathcal{P}_{t+1}(x^1,x^2)=&\mathcal{P}_t(x^1-1,x^2-1)\left[a_1+\dfrac{b_1}{t}+\dfrac{c_1}{t^2}\right]\nonumber\\
+&\mathcal{P}_t(x^1-1,x^2+1)\left[a_2+\dfrac{b_2}{t}+\dfrac{c_2}{t^2}\right]\nonumber\\
+&\mathcal{P}_t(x^1+1,x^2-1)\left[a_3+\dfrac{b_3}{t}+\dfrac{c_3}{t^2}\right]\nonumber\\
+&\mathcal{P}_t(x^1+1,x^2+1)\left[a_4+\dfrac{b_4}{t}+\dfrac{c_4}{t^2}\right]\label{jump}
\end{align}
where $\mathcal{P}_t(x^1,x^2)$ is the probability of having the displacement $x^1$ and $x^2$ at time $t$, and $a_i,b_i,c_i$ depend on $x^1$ and $x^2$ but none depend on time. We can approximate this process to a Fokker-Planck equation in two dimensions
\begin{align}
\frac{\partial\mathcal{P}}{\partial t}(x^1,x^2,t)&=-\dfrac{1}{t}\nabla\left(\sum_{i=1}^2\sum_{j=1}^2\alpha_i^j\gamma_i^jx^i\hat{x}^j\mathcal{P}(x^1,x^2,t)\right)\nonumber\\&+\dfrac{1}{2}\nabla^2\mathcal{P}(x^1,x^2,t). \label{FP2}
\end{align}
This is a first-order approximation to a large time limit, as obtained in \cite{dasilva2013nongaussian} to the original one-dimensional ERW. In that case, it can be seen that in the normal diffusion regime, and even at the transition, the process follows a central limit theorem \cite{coletti2016central}, thus making the Fokker-Planck approximation a good one in this regime, but not in the super-diffusive regime. So, at least to $\alpha_1^2=0$, the approximation Eq.(\ref{FP2}) is expected to be valid when both directions are in the normal diffusive regime.

Now, one can calculate the moments of this distribution. In particular, for the Ox-Cow case of section~\ref{section3}, we get for the first moment
\begin{equation}
\begin{matrix}
\dfrac{d}{dt}\langle x^1\rangle=\dfrac{\alpha_1^1}{t}\langle x^1\rangle,\\
\\
\dfrac{d}{dt}\langle x^2\rangle=\dfrac{\alpha_1^2\gamma}{t}\langle x^1\rangle+\dfrac{\alpha_2^2(1-\gamma)}{t}\langle x^2\rangle,
\end{matrix}
\end{equation}
which has the solution
\begin{equation}
\langle x^1\rangle=\langle x^1(t_0)\rangle\left(\dfrac{t}{t_0}\right)^{\alpha_1^1},
\end{equation}
\begin{align}
\langle x^2\rangle&=\left[\langle x^2(t_0)\rangle -\dfrac{\alpha_1^2\gamma\langle x^1(t_0)\rangle}{\alpha_1^1-\alpha_2^2(1-\gamma)}\right]\left(\dfrac{t}{t_0}\right)^{\alpha_2^2(1-\gamma)}\nonumber\\&+\dfrac{\alpha_1^2\gamma\langle x^1(t_0)\rangle}{\alpha_1^1-\alpha_2^2(1-\gamma)}\left(\dfrac{t}{t_0}\right)^{\alpha_1^1},
\end{align}
which has the same behavior as Eqs.(\ref{sol3}) and (\ref{sol4}) with respect to the power laws, despite different coefficients, which is an effect of the approximation of large time.
For the second moment, the equations are
\begin{equation}
\dfrac{d}{dt}\langle x^lx^k\rangle=\dfrac{1}{t}\sum_{i=1}^2\left(\alpha_i^k\gamma_i^k\langle x^ix^l\rangle+\alpha_i^l\gamma_i^l\langle x^ix^k\rangle\right)+\delta_{kl},\label{evosegundo}
\end{equation}
with $l,k=1,2$. This equation is easily solved in the Ox-Cow case, when $\alpha_1^2=0$:
\begin{align}
\langle \left(x^1(t)\right)^2\rangle&=\left[\langle x^1(t_0)x^1(t_0)\rangle-\dfrac{t_0}{1-2\alpha_1^1}\right]\left(\dfrac{t}{t_0}\right)^{2\alpha_1^1}\nonumber\\&+\dfrac{t}{1-2\alpha_1^1},
\end{align}
\begin{align}
\langle \left(x^2(t)\right)^2\rangle&=\left[\langle x^2(t_0)x^2(t_0)\rangle-\dfrac{t_0}{1-2\alpha_2^2(1-\gamma)}\right]\nonumber\\&\times\left(\dfrac{t}{t_0}\right)^{2\alpha_2^2(1-\gamma)}+\dfrac{t}{1-2\alpha_2^2(1-\gamma)},
\end{align}
and also exhibit the same power law behavior found in the discrete calculations (equation (\ref{asymp})). It must be stressed that in the Ox-Cow model, the Cow always walks as an ERW, since it is a decoupled direction.

Moreover, by solving Eq.(\ref{evosegundo}) for $\alpha_1^2\ne0$, one can find the following asymptotic regimes for the Ox diffusion: \emph{(a)} if $1/2<\alpha_1^1<(1-\gamma)\alpha_2^2$, the walk is super-diffusive, with $\langle(x_t^2)^2\rangle\sim t^{2(1-\gamma)\alpha_2^2}$, while \emph{(b)} if $1/2<(1-\gamma)\alpha_2^2<\alpha_1^1$, the walk is super-diffusive with $\langle(x_t^2)^2\rangle\sim t^{2\alpha_1^1}$; both regimes are in accordance with the mentioned conjecture from \cite{ferreira2010anomalous}. \emph{(c)} if $1/2=\alpha_1^1>(1-\gamma)\alpha_2^2$ or $1/2=(1-\gamma)\alpha_2^2>\alpha_1^1$, then the walk is marginally super-diffusive, with $\langle(x_t^2)^2\rangle\sim tln(t)$. These three regimes have already appeared in the original ERW, however, \emph{(d)} if $1/2=\alpha_1^1=(1-\gamma)\alpha_2^2$, the walk is also marginally super-diffusive, but faster, with $\langle(x_t^2)^2\rangle\sim t\left[ln(t)\right]^3$, and \emph{(e)} for $1/2<\alpha_1^1=(1-\gamma)\alpha_2^2$, the walk is slightly faster than the previous super-diffusion regimes, with $\langle(x_t^2)^2\rangle\sim t^{2\alpha_1^1}\left[ln(t)\right]^2$. \emph{(f)} the is normal diffusive otherwise. The phase diagrams of Fig.(\ref{figura2}) show these behavioral patterns.

It is noteworthy that the parameter $\alpha_1^2$ does not appear in the phase diagram, since effects of multiplicative constants are neglected in this kind of analysis, in which only the highest power is taken into account. Thus, the shape of Fig.(\ref{figura1}) can be understood as a consequence of the finite time of simulations, which is also an interesting point, since in real systems, the observation time might not be large enough to consider only the asymptotic result.

\section{Multi-Dimensional Model}

The model introduced in section \ref{section2} can be straightforwardly extended to an $N-$dimensional one by appropriately changing the rules. The walk, as in the beginning of section \ref{section4}, is provided by
\begin{equation}
	\overrightarrow{X_{t+1}}=\overrightarrow{X_t}+\overrightarrow{\sigma_{t+1}}
\end{equation}
where $\overrightarrow{X_t}=\left(X_t^1,\ldots,X_t^N\right)$ and $\overrightarrow{\sigma_t}=\left(\sigma_t^1,\ldots,\sigma_t^N\right)$. The extended rules are
\begin{enumerate}
	\item At time $t+1$, a time $t'$ in the set $\{1,\ldots,t\}$ is chosen with uniform probability.
	\item the probability $\mathcal{P}\left[\sigma_{t+1}^i=\sigma|\{\sigma_{t'}^1,\ldots,\sigma_{t'}^N\}\right]$ is provided by
		\begin{equation}
			\sum_{k=1}^{N}\dfrac{1}{2}\left[1+(2p_k^i-1)\sigma\sigma_{t'}^k\right]\gamma_k^i,
		\end{equation}
		where $0\le\gamma_k^i\le1$ and $\sum_{k=1}^{N}\gamma_k^i=1$;
	\item the first steps, as there is no past to follow, are provided with probability
		\begin{equation}
			\mathcal{P}\left[\sigma_{1}^i=\sigma\right]=\sum_{k=1}^{N}\dfrac{1}{2}\left[1+(2q_k^i-1)\sigma\right]\gamma_k^i;
		\end{equation}
\end{enumerate}
where I have already combined rules $1$ and $3$, inserting without loss of generality the analogous extension of Eq.(\ref{condprob2}) as rule $2$.

Now, it is possible to calculate all the results expressed in equations (\ref{eq9}-\ref{beta}) simply by changing the maximum index of the sums from $k=2$ to $k=N$. In addition, the results to higher moments, Eqs.(\ref{eq14}) and (\ref{rec}) still hold, but the number of entries of $\mathbb{M}_t$ is dimension-dependent. Nonetheless, the Fokker-Planck equation to the bidimensional model can also be generalized by changing the upper limits of the sums from $k=2$ to $k=N$ and changing the differential operators to their $N-$dimensional forms,

\begin{align}
\frac{\partial\mathcal{P}}{\partial t}(\overrightarrow{x},t)&=-\dfrac{1}{t}\nabla\left(\sum_{i=1}^N\sum_{j=1}^N\alpha_i^j\gamma_i^jx^i\hat{x}^j\mathcal{P}(\overrightarrow{x},t)\right)\nonumber\\&+\dfrac{1}{2}\nabla^2\mathcal{P}(\overrightarrow{x},t),\label{FP}
\end{align}
where $\{\hat{x}^j\}$ is an orthonormal basis of a hypercubic lattice where the walk takes place. It must be stressed that this equation is only a first-order approximation of long time behaviour in the continuous limit.

\section{Conclusions}
I have introduced a straightforward multi-dimensional generalization of the ERW in a way that takes into account the past motion in all directions as an influence on the next step. The recursion formula for the first moment of the displacement probability distribution has been explicitly calculated. The complexity in the treatment of the present model is due to the non-diagonal matrices involved in the solution for the higher-order moments recursion formulas (Eq.(\ref{rec})), which hamper long-term behavior analysis.

I have also highlighted two interesting regimes that occur in the \emph{Cow-and-Ox model} in section~\ref{section3}, the \emph{attractive} and the \emph{repulsive}, expressed by $(i.a)$ and $(i.c)$ of Eq.(\ref{regimei}), respectively. The former gives a hint on how this generalization can be used to model collective behavior, since this \emph{detective} regime may be viewed as a simple queue. Considering different walkers, instead of a single multi-dimensional walker, the coupled memory conects them all together, acting like an interaction force on a dynamic equation. 

The continuous process expressed by the Fokker-Planck equation (\ref{FP}) is only a first-order approximation of the discrete model, as in ref. \cite{schutz2004elephants}. Its higher-order corrections might be studied as in ref. \cite{dasilva2013nongaussian}. However, Eq.(\ref{FP}) is enough to describe the long-term behaviour of the first and second moments of the discrete model, therefore being useful to calculate diffusion behavior. With this equation, I found this model to exhibit two different anomalous diffusion regimes, not shown before by another ERW. Moreover, Eq.(\ref{FP}) can be reduced to that found in \cite{schutz2004elephants}, which in terms of the formalism I introduced can be viewed as a simple special case.

In future research, the model might be enriched with, for instance, the introduction of a \emph{stop} possibility, as in \cite{kumar2010memory}, and of \emph{position-dependent} coupling coefficients, while also discussing how crowd behavior might emerge from this microscopic coupling of different random walkers. The stop possibility has already been introduced in two dimensional models \cite{cressoni2013exact2d,dasilva2015522}, but the formalism has some differences. First, in those walks, unlike the coupling model, there are no steps taken simultaneously in both directions, which is basically what happens in the formalism developed here, and secondly, because of coupling, steps taken at a certain direction are not entirely set according the same direction, in fact, direction $i$ could stop by remembering a null step of direction $j$.
\appendix*

\begingroup
\allowdisplaybreaks
\begin{widetext}
\section{Bidimensional Jump Equation}

In order to derive the jump equation, it is necessary to introduce the complex characteristic function. Here I calculate the recurrence Eq.(\ref{reccomplex}) and then show the route to the jump equation (\ref{jump}). Let the walk be on a square lattice with orthonormal basis given by $\hat{x}^1=(1,0)$ and $\hat{x}^2=(0,1)$, so $\overrightarrow{x_t}=\left(x_t^1,x_t^2\right)$ and $\overrightarrow{\sigma_t}=\left(\sigma_t^1,\sigma_t^2\right)$. According to the definition $Q_t(\overrightarrow{k})=\langle e^{i\overrightarrow{k}\cdot\overrightarrow{x}_t}\rangle$ with $\overrightarrow{k}=(k^1,k^2)$, then
\begin{equation}
	Q_t(\overrightarrow{k})=\langle e^{ik^1x_t^1}\rangle\langle e^{ik^2x_t^2}\rangle.
\end{equation}
By evaluating the average $\langle e^{ik^ix_{t+1}^i}\rangle$
\begin{align}
	\langle e^{ik^ix_{t+1}^i}\rangle&=\langle e^{ik^ix_t^i}e^{ik^i\sigma_{t+1}^i}\rangle\nonumber\\&=\sum_{x_t^i}\sum_{\sigma_{t+1}^i}\mathcal{P}(x_t^i,\sigma_{t+1}^i)e^{ik^ix_t^i}e^{ik^i\sigma_{t+1}^i}\nonumber\\&=\sum_{x_t^i}\sum_{\sigma_{t+1}^i}\mathcal{P}(x_t^i)\mathcal{P}(\sigma_{t+1}^i|x_t^i)e^{ik^ix_t^i}e^{ik^i\sigma_{t+1}^i},
\end{align}
with Eq.(\ref{eq9}), we have
\begin{align}
		\langle e^{ik^ix_{t+1}^i}\rangle&=\sum_{x_t^i}\mathcal{P}(x_t^i)\sum_{\sigma_{t+1}^i}\left[\dfrac{1}{2}+\sigma_{t+1}^i\sum_{j=1}^{2}\dfrac{x_t^j\alpha_j^i\gamma_j^i}{2t}\right]e^{ik^ix_t^i}e^{ik^i\sigma_{t+1}^i}\nonumber\\&=\sum_{x_t^i}\mathcal{P}(x_t^i)e^{ik^ix_t^i}\left\{e^{ik^i}\left[\dfrac{1}{2}+\sigma_{t+1}^i\sum_{j=1}^{2}\dfrac{x_t^j\alpha_j^i\gamma_j^i}{2t}\right]+e^{-ik^i}\left[\dfrac{1}{2}-\sigma_{t+1}^i\sum_{j=1}^{2}\dfrac{x_t^j\alpha_j^i\gamma_j^i}{2t}\right]\right\}\nonumber\\&=\sum_{x_t^i}\mathcal{P}(x_t^i)e^{ik^ix_t^i}\left\{\cos(k^i)+i\sin(k^i)\sum_{j=1}^{2}\dfrac{x_t^j\alpha_j^i\gamma_j^i}{t}\right\}\nonumber\\&=\cos(k^i)\langle e^{ik^ix_t^i}\rangle+\sin(k^i)\sum_{x_t^i}\mathcal{P}(x_t^i) e^{ik^ix_t^i}\sum_{j=1}^{2}\dfrac{x_t^j\alpha_j^i\gamma_j^i}{t}.
\end{align}
The second term can be calculated as follows:
\begin{align}
	\sum_{x_t^i}\mathcal{P}(x_t^i)e^{ik^ix_{t}^i}\left[\dfrac{x_t^i\alpha_i^i\gamma_i^i}{t}+\sum_{j\ne i}\dfrac{x_t^j\alpha_j^i\gamma_j^i}{t}\right]&=	\sum_{x_t^i}\mathcal{P}(x_t^i)x_t^ie^{ik^ix_{t}^i}\dfrac{\alpha_i^i\gamma_i^i}{t}+\langle e^{ik^ix_{t}^i}\rangle\sum_{j\ne i}\dfrac{x_t^j\alpha_j^i\gamma_j^i}{t}\nonumber\\&=\dfrac{\alpha_i^i\gamma_i^i}{it}\dfrac{\partial}{\partial k^i}\left(\sum_{x_t^i}\mathcal{P}(x_t^i)e^{ik^ix_{t}^i}\right)+\langle e^{ik^ix_{t}^i}\rangle\sum_{j\ne i}\dfrac{x_t^j\alpha_j^i\gamma_j^i}{t}\nonumber\\&=\dfrac{\alpha_i^i\gamma_i^i}{it}\dfrac{\partial}{\partial k^i}\langle e^{ik^ix_{t}^i}\rangle+\langle e^{ik^ix_{t}^i}\rangle\sum_{j\ne i}\dfrac{x_t^j\alpha_j^i\gamma_j^i}{t},
\end{align}
and so,
\begin{align}
	\langle e^{ik^1x_{t+1}^1}\rangle&=\cos(k^1)\langle e^{ik^1x_t^1}\rangle+\sin(k^1)\dfrac{\alpha_1^1\gamma_1^1}{t}\dfrac{\partial}{\partial k^1}\langle e^{ik^1x_{t}^1}\rangle+i\sin(k^1)\langle e^{ik^1x_{t}^1}\rangle\dfrac{x_t^2\alpha_2^1\gamma_2^1}{t}\\
	\langle e^{ik^2x_{t+1}^2}\rangle&=\cos(k^2)\langle e^{ik^2x_t^2}\rangle+\sin(k^2)\dfrac{\alpha_2^2\gamma_2^2}{t}\dfrac{\partial}{\partial k^2}\langle e^{ik^2x_{t}^2}\rangle+i\sin(k^2)\langle e^{ik^2x_{t}^2}\rangle\dfrac{x_t^1\alpha_1^2\gamma_1^2}{t}.
\end{align}

Now, after multiplying $\langle e^{ik^1x_{t+1}^1}\rangle\langle e^{ik^2x_{t+1}^2}\rangle$ and rearranging the terms conveniently, we get
\begin{align}
Q_{t+1}(\overrightarrow{k})&=\cos(k^1)\cos(k^2)Q_t(\overrightarrow{k})\nonumber\\&+\cos(k^1)\sin(k^2)\sum_{i=1}^2\dfrac{\partial Q_t}{\partial k^i}\dfrac{\alpha_i^2\gamma_i^2}{t}\nonumber\\&+\cos(k^2)\sin(k^1)\sum_{i=1}^2\dfrac{\partial Q_t}{\partial k^i}\dfrac{\alpha_i^1\gamma_i^1}{t}\nonumber\\&+\sin(k^1)\sin(k^2)\sum_{i=1}^2\sum_{j=1}^2\dfrac{\partial^2 Q_t}{\partial k^i\partial k^j}\dfrac{\alpha_i^1\gamma_i^1\alpha_j^2\gamma_j^2}{t^2},\label{reccomplex2}
\end{align}
thus demonstrating equation (\ref{reccomplex}).

Now, we take the inverse Fourier transform,
\begin{equation}
	\mathcal{P}(\overrightarrow{x}_t)=\dfrac{1}{(2\pi)^2}\int Q_t(\overrightarrow{k})e^{-i\overrightarrow{k}\cdot\overrightarrow{x}_t}dk^1dk^2=\mathcal{F}^{-1}\left[Q_t(\overrightarrow{k})\right],\label{fourier}
\end{equation}
on the Eq.(\ref{reccomplex2}). Below, I evaluate each term of the calculation.

First, as stated in Eq.(\ref{fourier}),
\begin{equation}
	\mathcal{F}^{-1}\left[Q_{t+1}(\overrightarrow{k})\right]=\mathcal{P}(\overrightarrow{x}_{t+1})\equiv\mathcal{P}_{t+1}(x^1,x^2).\label{parte1}
\end{equation}
The next term is
\begin{align}
&\mathcal{F}^{-1}\left[\cos(k^1)\cos(k^2)Q_t(\overrightarrow{k})\right]\nonumber\\&=\dfrac{1}{(2\pi)^2}\int\cos(k^1)\cos(k^2)Q_t(\overrightarrow{k})e^{-i\overrightarrow{k}\cdot\overrightarrow{x}_t}dk^1dk^2,
\end{align}
and by representing the cosines with complex exponentials,

\begin{align}
	&\dfrac{1}{(2\pi)^2}\int\dfrac{Q_t(\overrightarrow{k})}{4}\left[e^{i(k^1+k^2)}+e^{i(k^1-k^2)}+e^{-i(k^1-k^2)}+e^{-i(k^1+k^2)}\right]e^{-ik^1x_t^1-ik^2x_t^2}dk^1dk^2\nonumber\\&=\dfrac{1}{4}\left\{\dfrac{1}{(2\pi)^2}\int Q_t(\overrightarrow{k})e^{-ik^1(x^1-1)-ik^2(x^2-1)}dk^1dk^2+\dfrac{1}{(2\pi)^2}\int Q_t(\overrightarrow{k})e^{-ik^1(x^1-1)-ik^2(x^2+1)}dk^1dk^2\right.\nonumber\\&\left.+\dfrac{1}{(2\pi)^2}\int Q_t(\overrightarrow{k})e^{-ik^1(x^1+1)-ik^2(x^2-1)}dk^1dk^2+\dfrac{1}{(2\pi)^2}\int Q_t(\overrightarrow{k})e^{-ik^1(x^1+1)-ik^2(x^2+1)}dk^1dk^2\right\}\nonumber\\&=\dfrac{1}{4}\left[\mathcal{P}_{t}(x^1-1,x^2-1)+\mathcal{P}_{t}(x^1-1,x^2+1)+\mathcal{P}_{t}(x^1+1,x^2-1)+\mathcal{P}_{t}(x^1+1,x^2+1)\right];\label{parte2}\\
	&\mathcal{F}^{-1}\left[\cos(k^1)\sin(k^2)\sum_{i=1}^2\dfrac{\partial Q_t}{\partial k^i}\dfrac{\alpha_i^2\gamma_i^2}{t}\right]\nonumber\\&=\dfrac{1}{(2\pi)^2}\int\cos(k^1)\sin(k^2)\sum_{i=1}^2\dfrac{\partial Q_t}{\partial k^i}\dfrac{\alpha_i^2\gamma_i^2}{t}e^{-i\overrightarrow{k}\cdot\overrightarrow{x}_t}dk^1dk^2\nonumber\\&=\dfrac{1}{(2\pi)^2}\int\dfrac{-i}{4}\left[e^{i(k^1+k^2)}-e^{i(k^1-k^2)}+e^{-i(k^1-k^2)}-e^{-i(k^1+k^2)}\right]e^{-ik^1x_t^1-ik^2x_t^2}\left(\dfrac{\partial Q_t}{\partial k^1}\dfrac{\alpha_1^2\gamma_1^2}{t}+\dfrac{\partial Q_t}{\partial k^2}\dfrac{\alpha_2^2\gamma_2^2}{t}\right)dk^1dk^2\nonumber\\&=\dfrac{1}{(2\pi)^2}\int\dfrac{-i}{4}\left[e^{-ik^1(x^1-1)-ik^2(x^2-1)}-e^{-ik^1(x^1-1)-ik^2(x^2+1)}+e^{-ik^1(x^1+1)-ik^2(x^2-1)}-e^{-ik^1(x^1+1)-ik^2(x^2+1)}\right]\nonumber\\&\times\left(\dfrac{\partial Q_t}{\partial k^1}\dfrac{\alpha_1^2\gamma_1^2}{t}+\dfrac{\partial Q_t}{\partial k^2}\dfrac{\alpha_2^2\gamma_2^2}{t}\right)dk^1dk^2\label{eqdomeio}\\&=\dfrac{1}{4}\dfrac{\alpha_1^2\gamma_1^2}{t}\dfrac{1}{(2\pi)^2}\left\{\int Q_t(\overrightarrow{k})(x^1-1)e^{-ik^1(x^1-1)-ik^2(x^2-1)}dk^1dk^2-\int Q_t(\overrightarrow{k})(x^1-1)e^{-ik^1(x^1-1)-ik^2(x^2+1)}dk^1dk^2\right.\nonumber\\&\left.+\int Q_t(\overrightarrow{k})(x^1+1)e^{-ik^1(x^1+1)-ik^2(x^2-1)}dk^1dk^2-\int Q_t(\overrightarrow{k})(x^1-1)e^{-ik^1(x^1+1)-ik^2(x^2+1)}dk^1dk^2\right\}\nonumber\\&+\dfrac{1}{4}\dfrac{\alpha_2^2\gamma_2^2}{t}\dfrac{1}{(2\pi)^2}\left\{\int Q_t(\overrightarrow{k})(x^2-1)e^{-ik^1(x^1-1)-ik^2(x^2-1)}dk^1dk^2-\int Q_t(\overrightarrow{k})(x^2+1)e^{-ik^1(x^1-1)-ik^2(x^2+1)}dk^1dk^2\right.\nonumber\\&\left.+\int Q_t(\overrightarrow{k})(x^2-1)e^{-ik^1(x^1+1)-ik^2(x^2-1)}dk^1dk^2-\int Q_t(\overrightarrow{k})(x^2+1)e^{-ik^1(x^1+1)-ik^2(x^2+1)}dk^1dk^2\right\}\label{eqdomeio2}\\&=\dfrac{\alpha_1^2\gamma_1^2}{4t}\left[(x^1-1)\mathcal{P}_t(x^1-1,x^2-1)-(x^1-1)\mathcal{P}_t(x^1-1,x^2+1)\right.\nonumber\\&\left.+(x^1+1)\mathcal{P}_t(x^1+1,x^2-1)-(x^1+1)\mathcal{P}_t(x^1+1,x^2+1)\right]\nonumber\\&+\dfrac{\alpha_2^2\gamma_2^2}{4t}\left[(x^2-1)\mathcal{P}_t(x^1-1,x^2-1)-(x^2+1)\mathcal{P}_t(x^1-1,x^2+1)\right.\nonumber\\&\left.+(x^2-1)\mathcal{P}_t(x^1+1,x^2-1)-(x^2+1)\mathcal{P}_t(x^1+1,x^2+1)\right].\label{parte3}
\end{align}

The next terms are evaluated similarly, and the results are
\begin{align}
	&\mathcal{F}^{-1}\left[\cos(k^2)\sin(k^1)\sum_{i=1}^2\dfrac{\partial Q_t}{\partial k^i}\dfrac{\alpha_i^1\gamma_i^1}{t}\right]\nonumber\\&=\dfrac{\alpha_1^1\gamma_1^1}{4t}\left[(x^1-1)\mathcal{P}_t(x^1-1,x^2-1)-(x^1+1)\mathcal{P}_t(x^1+1,x^2-1)\right.\nonumber\\&\left.+(x^1-1)\mathcal{P}_t(x^1-1,x^2+1)-(x^1+1)\mathcal{P}_t(x^1+1,x^2+1)\right]\nonumber\\&+\dfrac{\alpha_2^1\gamma_2^1}{4t}\left[(x^2-1)\mathcal{P}_t(x^1-1,x^2-1)-(x^2-1)\mathcal{P}_t(x^1+1,x^2-1)\right.\nonumber\\&\left.+(x^2+1)\mathcal{P}_t(x^1-1,x^2+1)-(x^2+1)\mathcal{P}_t(x^1+1,x^2+1)\right];\label{parte4}\\
	&\mathcal{F}^{-1}\left[\sin(k^1)\sin(k^2)\sum_{i=1,j=1}^2\dfrac{\partial^2 Q_t}{\partial k^i\partial k^j}\dfrac{\alpha_i^1\gamma_i^1\alpha_j^2\gamma_j^2}{t^2}\right]\nonumber\\&=\dfrac{\alpha_1^1\gamma_1^1\alpha_1^2\gamma_1^2}{4t^2}\left[(x^1-1)^2\mathcal{P}_t(x^1-1,x^2-1)-(x^1-1)^2\mathcal{P}_t(x^1-1,x^2+1)\right.\nonumber\\&\left.-(x^1+1)^2\mathcal{P}_t(x^1+1,x^2-1)+(x^1+1)^2\mathcal{P}_t(x^1+1,x^2+1)\right]\nonumber\\&+\dfrac{\alpha_1^1\gamma_1^1\alpha_2^2\gamma_2^2}{4t^2}\left[(x^1-1)(x^2-1)\mathcal{P}_t(x^1-1,x^2-1)-(x^1-1)(x^2+1)\mathcal{P}_t(x^1-1,x^2+1)\right.\nonumber\\&\left.-(x^1+1)(x^2-1)\mathcal{P}_t(x^1+1,x^2-1)+(x^1+1)(x^2+1)\mathcal{P}_t(x^1+1,x^2+1)\right]\nonumber\\&+\dfrac{\alpha_2^1\gamma_2^1\alpha_1^2\gamma_1^2}{4t^2}\left[(x^1-1)(x^2-1)\mathcal{P}_t(x^1-1,x^2-1)-(x^1-1)(x^2+1)\mathcal{P}_t(x^1-1,x^2+1)\right.\nonumber\\&\left.-(x^1+1)(x^2-1)\mathcal{P}_t(x^1+1,x^2-1)+(x^1+1)(x^2+1)\mathcal{P}_t(x^1+1,x^2+1)\right]\nonumber\\&+\dfrac{\alpha_2^1\gamma_2^1\alpha_2^2\gamma_2^2}{4t^2}\left[(x^2-1)^2\mathcal{P}_t(x^1-1,x^2-1)-(x^2+1)^2\mathcal{P}_t(x^1-1,x^2+1)\right.\nonumber\\&\left.-(x^2-1)^2\mathcal{P}_t(x^1+1,x^2-1)+(x^2+1)^2\mathcal{P}_t(x^1+1,x^2+1)\right].\label{parte5}
\end{align}

Now, by combining the results expressed in equations (\ref{parte1}), (\ref{parte2}), (\ref{parte3}), (\ref{parte4}) and (\ref{parte5}), we get

\begin{align}
\mathcal{P}_{t+1}(x^1,x^2)&=\mathcal{P}_{t}(x^1-1,x^2-1)\left[\dfrac{1}{4}+\dfrac{\alpha_1^2\gamma_1^2}{4t}(x^1-1)+\dfrac{\alpha_2^2\gamma_2^2}{4t}(x^2-1)+\dfrac{\alpha_1^1\gamma_1^1}{4t}(x^1-1)+\dfrac{\alpha_2^1\gamma_2^1}{4t}(x^2-1)\right.\nonumber\\&\left.+\dfrac{\alpha_1^1\gamma_1^1\alpha_1^2\gamma_1^2}{4t^2}(x^1-1)^2+\dfrac{\alpha_1^1\gamma_1^1\alpha_2^2\gamma_2^2}{4t^2}(x^1-1)(x^2-1)+\dfrac{\alpha_2^1\gamma_2^1\alpha_1^2\gamma_1^2}{4t^2}(x^1-1)(x^2-1)+\dfrac{\alpha_2^1\gamma_2^1\alpha_2^2\gamma_2^2}{4t^2}(x^2-1)^2\right]\nonumber\\&+\mathcal{P}_{t}(x^1-1,x^2+1)\left[\dfrac{1}{4}-\dfrac{\alpha_1^2\gamma_1^2}{4t}(x^1-1)-\dfrac{\alpha_2^2\gamma_2^2}{4t}(x^2+1)+\dfrac{\alpha_1^1\gamma_1^1}{4t}(x^1-1)+\dfrac{\alpha_2^1\gamma_2^1}{4t}(x^2+1)\right.\nonumber\\&\left.-\dfrac{\alpha_1^1\gamma_1^1\alpha_1^2\gamma_1^2}{4t^2}(x^1-1)^2-\dfrac{\alpha_1^1\gamma_1^1\alpha_2^2\gamma_2^2}{4t^2}(x^1-1)(x^2+1)-\dfrac{\alpha_2^1\gamma_2^1\alpha_1^2\gamma_1^2}{4t^2}(x^1-1)(x^2+1)-\dfrac{\alpha_2^1\gamma_2^1\alpha_2^2\gamma_2^2}{4t^2}(x^2+1)^2\right]\nonumber\\&+\mathcal{P}_{t}(x^1+1,x^2-1)\left[\dfrac{1}{4}+\dfrac{\alpha_1^2\gamma_1^2}{4t}(x^1+1)+\dfrac{\alpha_2^2\gamma_2^2}{4t}(x^2-1)-\dfrac{\alpha_1^1\gamma_1^1}{4t}(x^1+1)-\dfrac{\alpha_2^1\gamma_2^1}{4t}(x^2-1)\right.\nonumber\\&\left.-\dfrac{\alpha_1^1\gamma_1^1\alpha_1^2\gamma_1^2}{4t^2}(x^1+1)^2-\dfrac{\alpha_1^1\gamma_1^1\alpha_2^2\gamma_2^2}{4t^2}(x^1+1)(x^2-1)-\dfrac{\alpha_2^1\gamma_2^1\alpha_1^2\gamma_1^2}{4t^2}(x^1+1)(x^2-1)-\dfrac{\alpha_2^1\gamma_2^1\alpha_2^2\gamma_2^2}{4t^2}(x^2-1)^2\right]\nonumber\\&+\mathcal{P}_{t}(x^1+1,x^2+1)\left[\dfrac{1}{4}-\dfrac{\alpha_1^2\gamma_1^2}{4t}(x^1+1)-\dfrac{\alpha_2^2\gamma_2^2}{4t}(x^2+1)-\dfrac{\alpha_1^1\gamma_1^1}{4t}(x^1+1)-\dfrac{\alpha_2^1\gamma_2^1}{4t}(x^2+1)\right.\nonumber\\&\left.+\dfrac{\alpha_1^1\gamma_1^1\alpha_1^2\gamma_1^2}{4t^2}(x^1+1)^2+\dfrac{\alpha_1^1\gamma_1^1\alpha_2^2\gamma_2^2}{4t^2}(x^1+1)(x^2+1)+\dfrac{\alpha_2^1\gamma_2^1\alpha_1^2\gamma_1^2}{4t^2}(x^1+1)(x^2+1)+\dfrac{\alpha_2^1\gamma_2^1\alpha_2^2\gamma_2^2}{4t^2}(x^2+1)^2\right],
\end{align}
which is the complete form of the equation (\ref{jump}), which characterizes the described random walk as a jump process.

\end{widetext}

\begin{acknowledgments}
I would like to thank Professor D. O. Soares-Pinto and Professor L. P. Maia for the helpful discussions. This study was carried out with the support of CNPq, the National Council of Scientific and Technological Development - Brazil.
\end{acknowledgments}


%

\end{document}